# Broadband black phosphorus optical modulator in visible to mid-infrared spectral range

*Rui Zhang,[1] Yuxia Zhang,[1] Haohai Yu,[1*] Huaijin Zhang, [1**] Ruinong Yang,[2] Bingchao Yang,[2] Zhongyuan Liu,[2***] and Jiyang Wang[1]*

[1]State Key Laboratory of Crystal Materials and Institute of Crystal Materials, Shandong University, Jinan 250100, China

[2] State Key Laboratory of Metastable Materials Science and Technology, Yanshan University, Qinghuangdao 066004, China

*E-mail: haohaiyu@sdu.edu.cn

**E-mail: huaijinzhang@sdu.edu.cn

****E-mail: liuzy0319@yahoo.com

# Abstract

Black phosphorous (BP), a two-dimensional (2D) material, has a direct bandgap, which fills up the bandgap lacuna left by graphene topological insulators and transition-metal dichalcogenides because of its dependence on the layers and applied strains. Theoretically, the direct and tunable band gap indicates the broadband applications in optoelectronics with high efficiencies in the spectral range from visible to mid-infrared. Here, a BP broadband optical modulator is experimentally constructed and the passively modulated lasers at 639 nm (red), 1.06 μm (near-infrared) and 2.1 μm (mid-infrared) are realized by using the BP optical modulator as the saturable absorber in bulk lasers. The obtained results provide a promising alternative for rare broadband optical modulators and broaden the application range of BP in photonics.

# 1. Introduction

Two-dimensional (2D) layered crystals including graphene,[1-3] topological insulators,[4] and transition-metal dichalcogenides[5] are considered as potential next-generation electronic and optoelectronic devices. In band gap structures, single Dirac cones with zero band gaps exist in graphene[2] and on the surface states of topological insulators,[6] and the semiconductor transition-metal dichalcogenides have large band gap from 1 ev to 2eV. [7-10] Recently, black phosphorus (BP) re-attracts research efforts, since its direct band gap is tunable from 2 eV to 0.3 eV by layers and strains [11-15] and fills up the lacuna left by graphene, topological insulators and transition-metal dichalcogenides. Combined the high carrier mobility and strong intrinsic in-plane anisotropy[12, 14, 16, 17], BP is expected to have promising applications in optoelectronics especially in the infrared spectral range and electronics.[17-19] As representatives, the promising BP field-effect transistors and absorbers for solar cells were demonstrated[18, 20] and proposed, [21] respectively.

Beside the promising applications as transistors or absorbers, the nonlinear response of a 2D semiconductor to the applied field is also amazing. The possible nonlinear response properties are determined crystalline structure based on the Neumann principle. [22] In structure, bulk BP crystal belongs to the space group Cmca which indicates that the multilayered BP crystal has a center of inversion and has no second order nonlinearity including the piezoelectric, ferroelectric, Pockels, second harmonic generation, etc. For the third order nonlinearity, the saturable absorption of BP was studied in this year with the results that few-layered BP has nonlinear response in the visible to mid-infrared range with a large saturable intensity, defined as the optical intensity when the optical absorption coefficient is reduced to half of its original value. Associated with the direct band gap determining the efficient absorption and emission, the results of optical nonlinearities theoretically indicate that BP should be a potential broadband saturable absorber applied in the modulation in the visible to mid-infrared lasers with large pulse energy. However, up to now, only few BP applications were demonstrated in 1.5 to 1.9 μm fiber lasers [23-26] and there is no any result reported in the bulk lasers either. Here, we report the experimental realization of a BP pulse modulator and its applications in crystal lasers in the wavelength range from the visible to mid-infrared for the first time to our knowledge. The results widen the BP applying spectral ranges and experimentally identify that BP is a promising broadband optical modulator in optics beyond electronics.

## 2. Results and discussion

### 2.1 Analysis of the saturable absorption and characterization of BP broadband modulator

In fact, saturable absorption belongs to the third-order nonlinearity in optics and is generated by the imaginary part of the complex third-order susceptibility of polarization. Saturable absorbers are key devices for the generation of pulsed lasers with high peak power and large pulse energy by the modulation of the absorption (loss) in the laser cavity. For a semiconductor, the electrons in the valence band can be driven towards the conduction band by the interband transition under the irradiation of light with photon energy larger than the band gap and will subsequently return to the valence band by the thermalization process. However, under the irradiation by light with high intensity, the final electronic states in the valence and conduction bands will be fully occupied at the time shorter than the recombination lifetime, block the further absorption and demonstrate as saturable absorption. In the practical applications, the saturable absorbers with broadband response are desirable. Conventional semiconductor saturable absorbers including GaAs and semiconductor saturable absorber mirrors (SESAM) are wavelength sensitive and respond to light at a given wavelength band. [27] Aiming at the broadband application, graphene,[28, 29] topological insulator,[30] and well fabricated carbon nanotubes [31] and $MoS_2$ [32] were developed. Based on the recent studies, [23-26, 33] we found that multi-layered BP has a bandgap of 0.3 eV corresponding to the photon wavelength of 4.1 μm and saturable intensity over 261 $GW/cm^2$ (several orders larger in magnitude than the developed low dimensional saturable absorbers), which theoretically indicates that BP can be used as a broadband optical modulator for the generation of large laser pulses with the wavelength shorter than 4.1 μm.

The mechanically exfoliated BP flake was investigated by Raman spectrometry. The Raman shifting stimulated by a laser with the wavelength of 633 nm is presented in **Figure 1a**. From this figure, we observe that the shifted peaks appear at 369.2 $cm^{-1}$, 439.0 $cm^{-1}$ and 467.5 $cm^{-1}$, corresponding to one out-of-plane mode $A_g^1$ and two in-plane vibration modes $B_{2g}$ and $A_g^2$, respectively.[34, 35] Because Raman spectra of BP are dependent on the layers,[36] the position of the Raman shifted peaks have negligible difference which agreed well with the previous reported samples with the layers about

30.[33] With atomic force microcopy (AFM), the morphology of prepared BP sample was studied as shown in **Figure 1b** and the thickness of the prepared samples was confirmed. In **Figure 1b**, we can see that BP distribute in platelets with the dimensions of around 2 μm×4 μm on the quartz glass wafer. The height difference between the substrate and target BP sections is exhibited in **Figure 1c** and is mainly from 25 to 30nm. Assuming single layer BP with a thickness of ~0.6 nm,[37] the prepared sample is estimated to have 40-50 layers, which means that the bandgap of the BP sample is about 0.3 eV and can respond to the photons with the wavelength shorter than 4.1 μm. In order to identify the potential applications of BP in the visible to mid-infrared spectral range, the absorption spectrum of the prepared BP sample on quartz glass was measured by using a V-570 JASCO UV/VIS/NIR spectrophotometer and is exhibited in **Figure 1d**. For comparison, the spectrum of a quartz glass substrate is also presented in **Figure 1d**.  From this figure, we can find that the absorption of BP decreases with the increase of wavelength and has absorption in the range from the visible to mid-infrared. The absorption results indicate that prepared BP sample can be used as a broadband saturable absorber provided that the photon energy is larger than 0.5 eV corresponding to the wavelength of 2500 nm.

**2.2 BP optical modulation visible to mid-infrared lasers**

Using the prepared BP sample as saturable absorbers for the modulation of laser operation, passively Q-switched lasers at wavelengths of 639 nm (red), 1.06 μm (near-infrared) and 2.1 μm (mid-infrared) were realized with Pr:GdLiF$_4$, Nd:GdVO$_4$ and Tm:Ho:Y$_3$Ga$_5$O$_{12}$ (Tm:Ho:YGG) crystals as gain materials. A detailed description of the laser experimental configurations for different operating wavelengths is shown in the *Experimental Section*. The pulse width and repetition rate of the passively Q-switched lasers with different wavelengths were recorded with a digital oscilloscope and a photodiode detector.

We present the average output power and repetition rate with the increase of incident pump power in **Figure 2a-c**. From these figures, we find that the output power and repetition rate of the all lasers increased with incident pump power, which agrees well with the passive Q-switching theory. [38] The achieved maximum average output power at the wavelength of 639 nm, 1.06 μm and 2.1 μm was 18 mW, 22 mW and 27 mW, respectively, with the respective threshold of 0.66 W, 0.33 W and 0.97 W. The maximum repetition rate for the three Q-switched lasers is 172 kHz, 312 kHz, and 122 kHz, respectively. With the average output power and repetition rate, the pulse energy can be calculated and shown in **Figure 2d-f** for different wavelengths. The maximum pulse energy is 104 nJ (639 nm), 70.4 nJ (1.06 μm) and 221 nJ (2.1 μm). The difference of the present pulse energy is induced by the dispersion of BP absorption and the different emission cross-sections of laser crystals used, since the large absorption of BP sample and a small emission cross-section of the laser crystal would be possible to generate the large pulse energy.[38] The variation of pulse width of the passively Q-switched lasers at different wavelength was also demonstrated in **Figure 2d-f**, with the shortest pulse width of 189 ns, 495 ns and 636 ns, respectively, With the pulse energy and pulse width, the peak power was calculated with the maximum values of 552 mW, 142 mW and 348mW for the wavelength of 639 nm, 1.06 μm and 2.1 μm, respectively.

**Figure 3a-c** present the typical pulses of the three passively Q-switched lasers, respectively, and the laser spectrum of the different lasers are shown in **Figure 3d-f**. It should be noted that there is no thermal damage on BP samples observed and the output power was stable during laser operation, which indicates that the pulse laser performance can be further improved by optimizing the quality and thickness of BP sample and the laser design. Besides the first applications in broadband bulk lasers and much more broad application range from visible (639 nm) to mid-infrared (2.1 μm), the present results also have some advantages in the aspects of pulse width and peak powers, an order shorter and larger in magnitude, respectively, compared with the recent reported BP Q-switched fiber lasers at the wavelength from 1.5 μm to 1.9 μm, [23]

## 3. Conclusion

A broadband BP optical modulator was constructed by taking advantages of the narrow and direct band gap and saturable absorption properties. With the developed BP sample, the passively Q-switched lasers at the wavelengths of 639 nm, 1.06 μm and 2.1 μm were realized, which widely expand the spectral applications of BP and identify that BP is a universal saturable absorber for lasers at the wavelength from the visible to mid-infrared. We also propose that the pulsed laser performance could be further improved by optimizing the BP sample and laser cavity designing. These results provide an alternative for the universal optical modulators and should be helpful for the further applications of BP beyond the present optoelectronics and electronics.

## 4. Experimental Section:

*Black phosphorus sample fabrication:*. Black phosphorus bulk crystal was synthesized from red phosphorous (99.999% purity, Alpha Asear) under the high temperature of 1000 ℃ and the high pressure of 2 GPa. The BP flakes were mechanically exfoliated from the synthesized BP crystal by using scotch tape, and they were transferred onto a commercial optical-grade far-ultraviolet quartz wafer (25mm diameter, 1mm thick) by pressing the tape surface against the wafer and then peeling it off slowly.

*Q-switched laser configuration:* The pump sources for Q-switched lasers experiments were laser diodes. All laser resonators consist of two mirrors. With focusing systems, pumping light was delivered through the front input mirror, antireflection (AR) coated for the pump wavelength and highly reflective (HR) for the laser wavelength. The laser crystals had polished and uncoated surfaces and were mounted in copper holders with circulating cool water. The prepared BP sample was inserted into the cavity near the output coupler. For the red laser, the pump source was a blue laser diode with the wavelength of 442 nm. A lens with the focal length of 25mm is used for focusing pumping light. The resonator is a plane-concave cavity. The front mirror was plane and the output coupler was a concave mirror with a 50 mm radius and transmission of 1.8% at 639 nm. The length between the front and output mirrors was optimized to be 45 mm. A Pr:GdLiF$_4$ crystal with dimensions of 2.5 mm ×2.0 mm ×6.9 mm (a×c×a) was used as the laser material. The Pr$^{3+}$ ions doping concentration in Pr:GdLiF$_4$ was 1.01at.%.

In the laser at the wavelength of 1.06 μm, a fiber-coupled laser diode at a wavelength of 808 nm was employed as the pump source with a numerical aperture of 0.22 and a 200 μm core diameter. A Nd:GdVO$_4$ laser crystal with a Nd$^{3+}$ doping concentration of 0.5 at.% had dimensions of 3 mm×3 mm×8 mm(a×b×c). The resonator was a plane-concave cavity. The front mirror was plane and the output mirror had a radius of curvature of 100 mm with a transmission of 20 % at 1.06μm. The length between the front and output mirrors was optimized to be 94 mm.

For the 2.1 μm laser, the pump source was a fiber-coupled diode at a wavelength of 795 nm, with a numerical aperture of 0.22 and a 200 μm core diameter. The laser crystal was Ho:Tm:YGG with dimensions of 3 mm×3 mm×4 mm cut along the <111> direction. The $Tm^{3+}$ and $Ho^{3+}$ doping concentrations were 5 at.% and 0.36 at.%, respectively. A concave-plane cavity was used for the Q-switched laser. The front mirror had a 200 mm radius of curvature and the output mirror was plane with transmission of 5 % at 2.1 μm. The length between the front and output mirrors was 24 mm.

## Acknowledgements

The National Natural Science Foundation of China (Nos. 51422205 and 51272131), the Natural Science Foundation for Distinguished Young Scholars of Shandong Province (2014JQE27019) and Taishan Scholar Foundation of Shandong Province, China.

**Figures:**

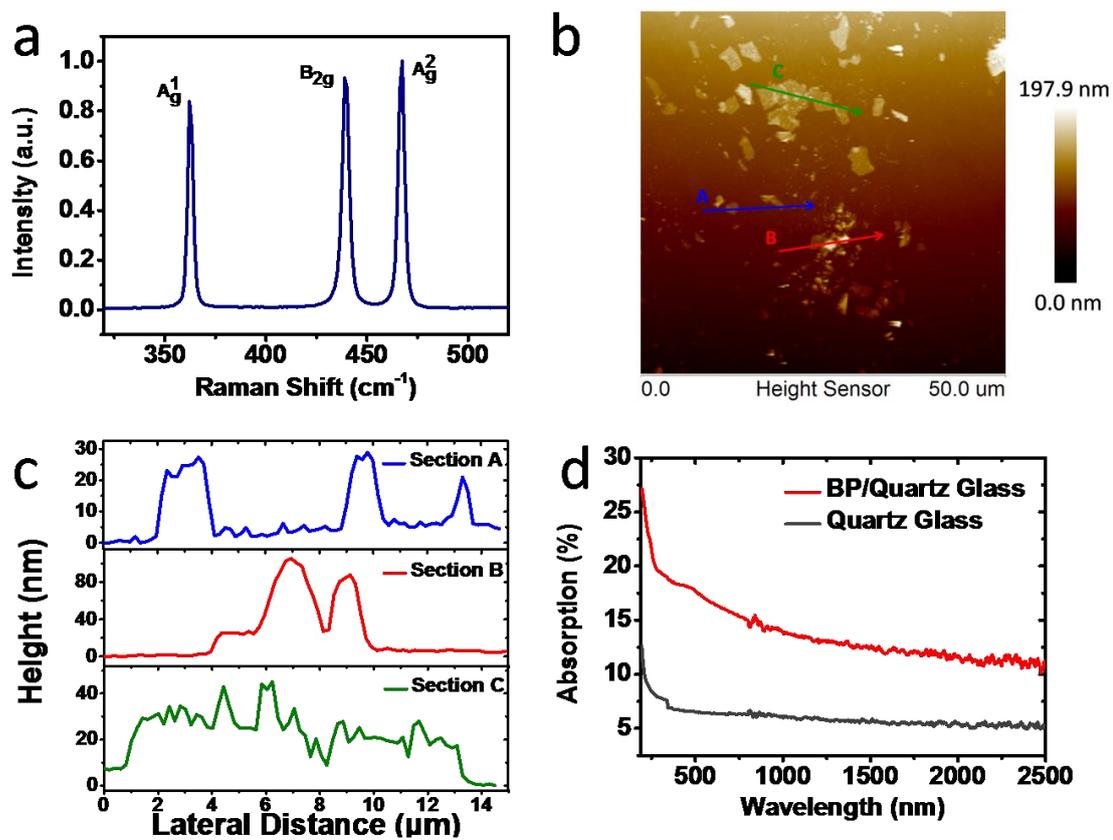

**Figure 1.** a) Raman spectrum of BP sample. b) Morphology of BP measured with atomic force microcopy. c) Height Profiles of the three sections marked in AFM image. d). Absorption spectrum of BP sample and substrate.

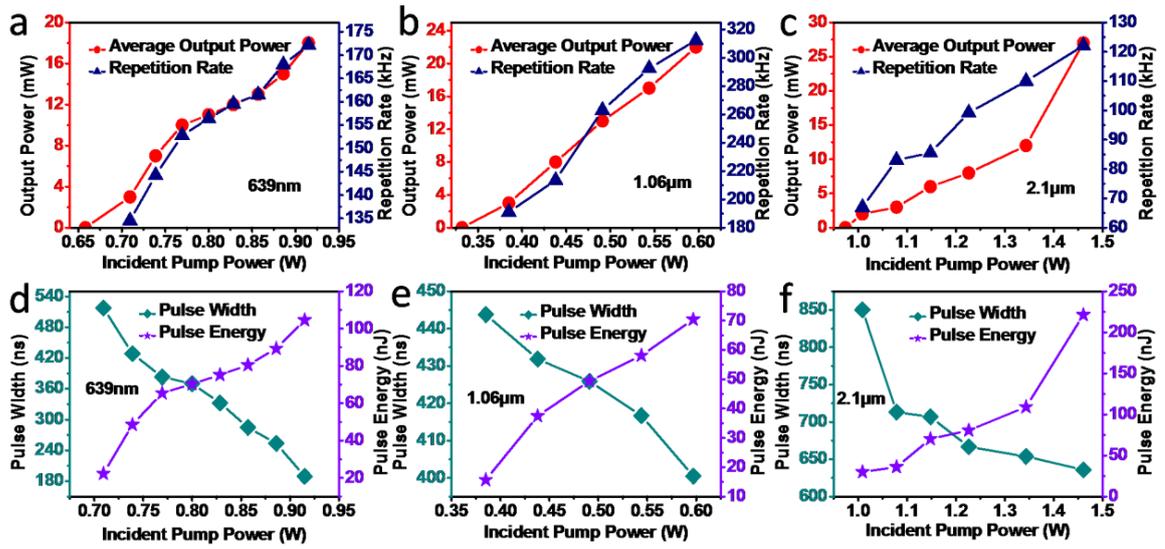

**Figure 2.** Passively Q-switched lasers performances. a-c) Average output power and repetition rate of passively Q-switched Pr:GdLiF$_4$, Nd:GdVO$_4$ and Tm:Ho:YGG laser at 639 nm, 1.06 μm and 2.1 μm, respectively. d-f) Pulse width and pulse energy of passively Q-switched Pr:GdLiF$_4$, Nd:GdVO$_4$ and Tm:Ho:YGG laser at 639 nm, 1.06 μm and 2.1 μm, respectively.

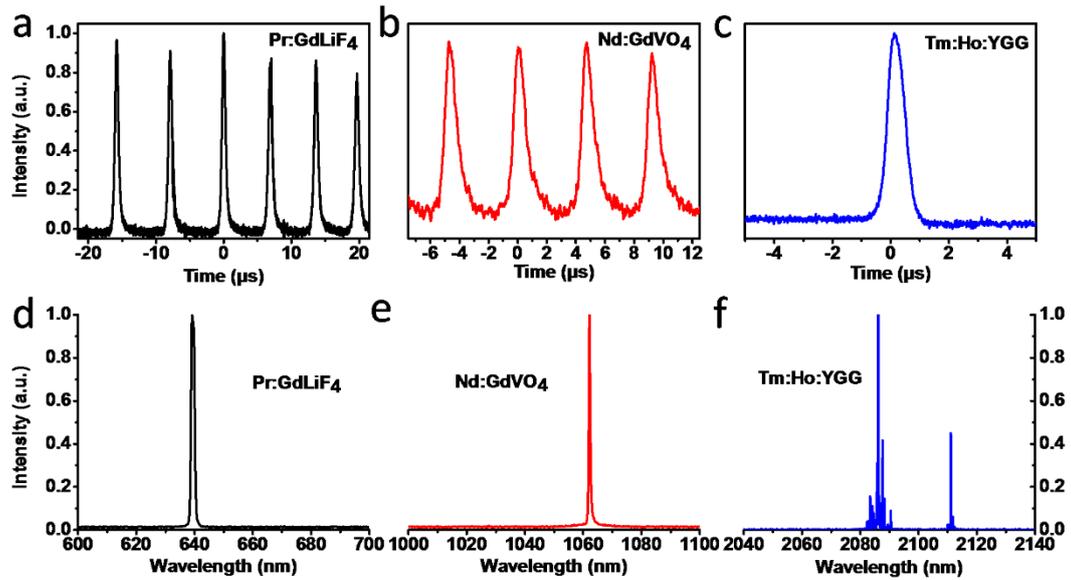

**Figure 3.** Pulses and laser spectra of passively Q-switched lasers. a-c) Passively Q-switched Pr:GdLiF$_4$, Nd:GdVO$_4$ and Tm:Ho:YGG laser with the pulse width of 189ns, 495ns and 636ns, respectively. d-f) Passively Q-switched Pr:GdLiF$_4$, Nd:GdVO$_4$ and Tm:Ho:YGG laser spectra at centre wavelength of 639 nm, 1.06 μm and 2.1 μm, respectively.